\documentclass[prc,aps,twocolumn,floats,floatfix,showpacs,superscriptaddress,nofootinbib]{revtex4}
\usepackage{amsmath}
\usepackage{amsfonts}
\usepackage{graphicx}

\begin{document}

\title{Influence of complex configurations on properties of pygmy dipole resonance
in neutron-rich Ca isotopes}

\author{N. N. Arsenyev}
\affiliation{Bogoliubov Laboratory of Theoretical Physics,
             Joint Institute for Nuclear Research,
             141980 Dubna, Moscow region, Russia}
\author{A. P. Severyukhin}
\affiliation{Bogoliubov Laboratory of Theoretical Physics,
             Joint Institute for Nuclear Research,
             141980 Dubna, Moscow region, Russia}
\affiliation{Dubna State University,
             141982 Dubna, Moscow region, Russia}
\author{V. V. Voronov}
\affiliation{Bogoliubov Laboratory of Theoretical Physics,
             Joint Institute for Nuclear Research,
             141980 Dubna, Moscow region, Russia}
\author{Nguyen Van Giai}
\affiliation{Institut de Physique Nucl\'eaire, CNRS-IN2P3 and
             Univ. Paris-Sud, 91405 Orsay, France}

\begin{abstract}
Starting from the quasiparticle random phase approximation based
on the Skyrme interaction SLy5, we study the effects of
phonon-phonon coupling~(PPC) on the low-energy electric dipole
response in $^{40-58}$Ca. Using the same set of parameters we
describe available experimental data for $^{40,44,48}$Ca and give
prediction for $^{50-58}$Ca. The inclusion of the PPC results in
the formation of low-energy $1^-$ states. There is an impact of
the PPC effect on low-energy $E1$~strength of $^{40,44,48}$Ca. The
PPC effect on the electric dipole polarizability is discussed. We
predict a strong increase of the summed $E1$~strength below
10~MeV, with increasing neutron number from $^{48}$Ca till
$^{58}$Ca.
\end{abstract}


\date{\today}

\maketitle
%
%
\section{Introduction}

Collective dipole excitations are a common phenomenon of finite
fermion systems. In atomic nuclei they can arise, for instance,
from out-of-phase oscillations of the proton and neutron
``fluids'' giving the well known giant dipole
resonance~(GDR)~\cite{Migdal44}. Systematic investigations
established it to be a global feature of nuclei from the very
light to the heaviest nuclei~\cite{Berman75,Dietrich88}. In recent
years, the interest is more focused on the low-lying dipole
strength, that is located below the GDR energies. The
concentration of the $E1$~strength around the particle separation
energy is commonly called the pygmy dipole resonance~(PDR) because
of its weak strength in comparison with the GDR, which dominates
the $E1$~strength distribution in nuclei~\cite{Savran13}. In
analogy to the GDR, the PDR has been interpreted as a collective
oscillation of the neutron skin with respect to a $N{\approx}Z$
inert core (see Ref.~\cite{Paar07} and references therein). The
total sum of the measured energy-weighted sum rule of such $E1$
distributions is less than 1-2{\%} of the Thomas-Reiche-Kuhn~(TRK)
sum rule value for stable nuclei and less than 5-6{\%} for
unstable neutron-rich nuclei~\cite{Savran13}. Recent theoretical
calculations indicate that such a low-energy peak is a common
property of neutron-rich nuclei lying in different mass
regions~\cite{Inakura11,Ebata14}. The occurrence of non-negligible
low-lying $E1$-strength can influence the radiative neutron
capture cross section by orders of magnitude and, consequently,
also the rate of the astrophysical $r$-process
nucleosynthesis~\cite{Arnould07}. The PDR study is expected to
provide information on the symmetry energy term of the nuclear
equation of state~\cite{Brown00,Horowitz01}.

The strong proton shell closure at $Z=20$ and the already good
experimental knowledge of the chain of calcium isotopes
makes~\cite{Hartmann00,Hartmann04,Isaak11,Derya14} calcium an
attractive clement for a PDR study. Indeed, indications of a trend
for increasing low-energy dipole strength with increasing mass can
be observed in the dipole excitation functions (above neutron
separation energy) in the stable Ca
isotopes~\cite{Hartmann00,Hartmann04}. The results were generally
consistent with the theoretical prediction regarding the shifting
of dipole strength to lower energies, see, e.g.,
Refs.~\cite{Inakura11,Soloviev78,Terasaki06,Egorova16}. Moreover,
recent experimental studies indicate $N=32$ as a new magic number
in Ca isotopes due to the high energy of the first $2^{+}$ state
in $^{52}$Ca~\cite{Gade06} and the trend obtained for the
two-neutron separation energies~\cite{Wienholtz13}. The first
experimental spectroscopic study of low-lying states in $^{54}$Ca
have been performed at RIKEN~\cite{Steppenbeck13}: the $2^{+}_{1}$
energy in $^{54}$Ca was found to be only $\sim500$ keV below that
in $^{52}$Ca, suggesting a $N=34$ new shell closure. Finally, we
note that new progress in the production of neutron-rich Ca
isotopes can be expected at the NSCL at Michigan State
University~\cite{Tarasov09}. Future measurements of excited states
and masses for the neighboring Ca isotopes will further enhance
our understanding of nuclear states in very neutron-rich systems.
Thus, the spectroscopy of neutron-rich calcium isotopes provides a
valuable information, with important tests of theoretical
calculations.

A powerful microscopic approach is the quasiparticle-phonon model
(QPM)~\cite{Soloviev92}. Its ability for the describing the
low-energy nuclear spectroscopy has recently been reviewed in
Ref.~\cite{LoIudice12}. The model Hamiltonian is diagonalized in a
space spanned by states composed of one, two and three phonons
which are generated in quasiparticle random phase approximation
(QRPA)~\cite{Grinberg94,Ponomarev98}. The separable form of the
residual interaction is the practical advantage of the QPM which
allows one to perform the structure calculations in large
configurational spaces. The mean field is modelized by a
Woods-Saxon potential well. These are the basic ingredients of the
QPM. The single-particle energies and ground-state properties in
general are critical quantities for extrapolations of QRPA and QPM
calculations into unknown mass regions. A special emphasis on a
reliable description of the mean-field part, reproducing as
closely as possible the ground-state properties of nuclei along an
isotopic chain has been done in~\cite{Tsoneva04,Tsoneva08}. This
is achieved by solving the ground-state problem in a
semi-microscopic approach based on a Skyrme energy density
functional (EDF)~\cite{Tsoneva04,Tsoneva08}. The EDF$+$QPM
calculations have been applied for calculating the low-energy
dipole strength~\cite{Tsoneva04,Tsoneva08}, as well as used for
astrophysical applications~\cite{Tsoneva15}. The results indicate
that the radiative capture cross sections are underestimated by a
factor of about two by the QRPA for the $N=50$ nuclei. This
sensitivity is the cause of the importance of the multiphonon
coupling and of the relevance of the EDF$+$QPM approach for
astrophysical applications.

The QRPA with a self-consistent mean-field derived from Skyrme EDF
is one of the most successful methods for studying the low-energy
dipole strength, see e.g.,
\cite{Sarchi04,Terasaki06,Avdeenkov11,Arsenyev12,Repko13}. Such an
approach describes the properties of the low-lying states less
accurately than more phenomenological ones, but the results are in
a reasonable agreement with experimental data. On the other hand,
due to the anharmonicity of vibrations there is a coupling between
one-phonon and more complex states~\cite{Soloviev92,LoIudice12}.
The main difficulty is that the complexity of calculations beyond
standard QRPA increases rapidly with the size of the configuration
space, so that one has to work within limited spaces. Using a
finite-rank separable
approximation~(FRSA)~\cite{Giai98,Severyukhin02,Severyukhin04,Severyukhin08}
for the residual interaction one can overcome this difficulty.
Alternative schemes to factorize the residual Skyrme interaction
have also been considered in
Refs.~\cite{Suzuki81,Sarriguren99,Nesterenko02}. The FRSA was thus
used to study the electric low-energy excitations and giant
resonances within and beyond the
QRPA~\cite{Severyukhin04,Severyukhin08,Severyukhin09,Severyukhin12}.
In this paper our approach applied for PDR features of
neutron-rich nuclei. We will give an illustration of our approach
for $^{48}$Ca with closed neutron shell in comparison to the
$N=30$ isotope $^{50}$Ca. Preliminary results of our studies for
neutron-rich Sn isotopes are reported in
Refs.~\cite{Arsenyev12,Arsenyev15,Arsenyev16}.

The paper is organized as follows. A brief summary of the
formalism including the effects of the phonon-phonon coupling is
given in Sec. II. Some details about the numerical calculations
are presented in Sec. III, while in Sec. IV, results are analyzed
and compared with available experimental data. Finally, our
conclusions are laid in the last section.

%
%
\section{The FRSA model}

The starting point of the method is the Hartree-Fock~(HF)-BCS
calculation~\cite{Ring80} of the ground state based on Skyrme
interactions. Spherical symmetry is imposed on the quasiparticle
wave functions. The continuous part of the single-particle
spectrum is discretized by diagonalizing the Skyrme HF Hamiltonian
on a harmonic oscillator basis. In the particle-hole (p-h) channel
we use the Skyrme interaction with the tensor components and their
inclusion leads to the modification of the spin-orbit
potential~\cite{Stancu77,Lesinski07}. The pairing correlations are
generated by a density-dependent zero-range force
\begin{equation}
V_{pair}({\bf r}_1,{\bf r}_2)=V_{0}\left[1-\eta\left(\frac{\rho
\left(r_{1}\right)}{\rho_{0}}\right)^{\gamma}\right]
\delta\left({\bf r}_{1}-{\bf r}_{2}\right), \label{pair}
\end{equation}
where $\rho\left(r_{1}\right)$ is the particle density in
coordinate space, $\rho _{0}$ being the nuclear matter saturation
density; $\gamma$, $\eta$, and $V_{0}$ are adjusted parameters.
The parameters are determined by adjusting the empirical odd-even
mass differences of the nuclei in the region under study.

To build the QRPA equations on the basis of HF-BCS quasiparticle
states, the residual interaction is consistently derived from the
Skyrme force in the p-h channel and from the zero-range pairing
force in the particle-particle (p-p) channel~\cite{Terasaki05}.
The FRSA for the residual interaction enables us to find QRPA
eigenvalues as the roots of a relatively simple secular
equation~\cite{Giai98,Severyukhin08}. The cut-off of the
discretized continuous part of the single-particle spectra is at
the energy of 100~MeV. This is sufficient to exhaust practically
all the sum rules and, in particular, the TRK sum rule with the
enhancement factor $\kappa$ for the EDF. The FRSA is discussed in
more details in
Refs.~\cite{Severyukhin08,Severyukhin12,Severyukhin14}.

To take into account the effects of the phonon-phonon
coupling~(PPC) we follow the basic QPM
ideas~\cite{Soloviev92,LoIudice12}. We construct the wave
functions from a linear combination of one-phonon and two-phonon
configurations
\begin{eqnarray}
\Psi_\nu(\lambda\mu)=\left(\sum\limits_{i}R_{i}(\lambda\nu)Q_{\lambda\mu
i}^{+}\right.\nonumber\\
\left.+\sum\limits_{\lambda_{1}i_{1}\lambda_{2}i_{2}}P_{\lambda_{2}i_{2}}^{\lambda_{1}i_{1}}
(\lambda\nu)\left[Q_{\lambda_{1}\mu_{1}i_{1}}^{+}Q_{\lambda_{2}\mu_{2}i_{2}}^{+}\right]_{\lambda\mu}\right)|0\rangle,
\label{wf2ph}
\end{eqnarray}
where $\lambda$ denotes the total angular momentum and $\mu$ its
z-projection in the laboratory system. The ground state is the
QRPA phonon vacuum $|0\rangle$. The unknown amplitudes
$R_i(\lambda\nu)$ and
$P_{\lambda_{2}i_{2}}^{\lambda_{1}i_{1}}(\lambda\nu)$ are
determined from the variational principle, which leads to a set of
linear equations~\cite{Severyukhin04,Severyukhin12}
\begin{eqnarray}
(\omega_{\lambda i}-E_{\nu})R_{i}(\lambda\nu )
\nonumber\\
+\sum\limits_{\lambda_{1}i_{1}\lambda_{2}i_{2}}U_{\lambda_{2}i_{2}}^{\lambda_{1}i_{1}}
(\lambda i)P_{\lambda_{2}i_{2}}^{\lambda_{1}i_{1}}(\lambda\nu)=0,
\label{2pheq1}
\end{eqnarray}
\begin{eqnarray}
2(\omega_{\lambda_{1}i_{1}}+\omega_{\lambda_{2}i_{2}}-E_{\nu})
P_{\lambda_{2}i_{2}}^{\lambda_{1}i_{1}}(\lambda\nu)\nonumber\\
+\sum\limits_{i}U_{\lambda_{2}i_{2}}^{\lambda_{1}i_{1}}(\lambda
i)R_{i}(\lambda\nu)=0.
\label{2pheq2}
\end{eqnarray}

The rank of the set of linear equations~(\ref{2pheq1}) and
(\ref{2pheq2}) is equal to the number of one- and two-phonon
configurations included in the wave function~(\ref{wf2ph}). Its
solution requires to compute the Hamiltonian matrix elements
coupling one- and two-phonon configurations:
\begin{equation}
U_{\lambda_{2}i_{2}}^{\lambda_{1}i_{1}}(\lambda i)=\langle0|
Q_{\lambda i} H
\left[Q_{\lambda_{1}i_{1}}^{+}Q_{\lambda_{2}i_{2}}^{+}\right]_{\lambda}|0
\rangle.
\label{2phU}
\end{equation}
Equations~(\ref{2pheq1}) and (\ref{2pheq2}) have the same form as
the QPM equations~\cite{Soloviev92,LoIudice12}, where the
single-particle spectrum and the residual interaction are derived
from the same Skyrme EDF.

%
%
\section{Details of calculations}

\begin{figure}[t!]
\includegraphics[width=1.0\columnwidth]{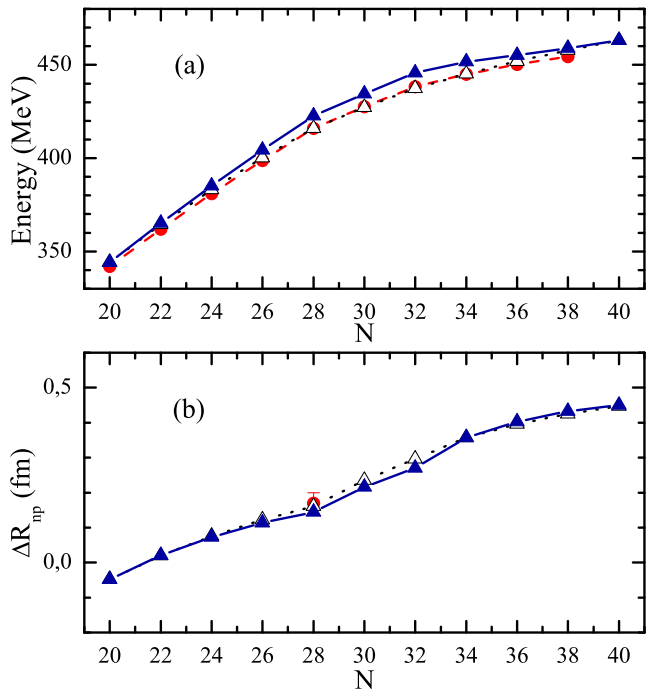}
\caption{(Color online) (a) Binding energies of the even-even Ca
isotopes as a function of neutron number, compared with experiment
and extrapolated energies (filled circles) from the AME2012 atomic
mass evaluation~\cite{Wang12}. Results of the calculations within
the HF-BCS with the SLy5 EDF (open triangles) and with SLy5{+}T
(filled triangles). (b) The neutron skin thickness~$\Delta R_{np}$
of the Ca isotopes calculated within the HF-BCS approach with the
SLy5 EDF (open triangles) and with SLy5{+}T (filled triangles).
Experimental data of the neutron skin thickness are taken from
Ref.~\cite{Birkhan16} (filled circles).} \label{CaBERpn}
\end{figure}
We apply the above approach to study the influence of the PPC on
the strength $E1$~distributions of the neutron-rich Ca isotopes.
We use the Skyrme interactions SLy5~\cite{Chabanat98} and
SLy5{+}T~\cite{Colo07} in the p-h channel. The parameters of the
force SLy5 have been adjusted to reproduce nuclear matter
properties, as well as nuclear charge radii, binding energies of
doubly magic nuclei. The force SLy5{+}T involves the tensor terms
added without refitting of the parameters of the central
interaction (the tensor interaction parameters are
$\alpha_T=-170$~MeV{}fm$^5$ and $\beta_T=100$~MeV{}fm$^5$). These
parametrizations describe correctly the binding energies of
even-even Ca isotopes. This is illustrated in
Fig.~\ref{CaBERpn}(a), where the calculated binding energies for
$^{40-60}$Ca and the experimental and extrapolated
data~\cite{Wang12} are shown. The agreement between the SLy5
results and data is reasonable, the deviations being less than
1{\%}. On the other hand, comparing SLy5 and SLy5{+}T results
shows that the maximum relative difference between the binding
energies is about 2{\%} for the case of $^{52}$Ca. In the case of
the SLy5{+}T EDF, these deviations are connected with the central
Skyrme parameters which have not been refitted after including the
tensor terms~\cite{Colo07}.

In Fig.~\ref{CaBERpn}(b), we show the neutron skin thickness of Ca
isotopes as a function of the neutron number. The neutron skin
thickness~$\Delta R_{np}$ is defined as
\begin{equation}
\Delta
R_{np}{=}\sqrt{\mathstrut{\langle{r^2}\rangle}_{n\vphantom{p}}}
-\sqrt{\mathstrut{\langle{r^2}\rangle}_{p}}. \label{rnp}
\end{equation}
As can be seen from Fig.~\ref{CaBERpn}(b), the proton-neutron rms
differences become larger when the neutron number increases. The
same evolution is obtained with other Skyrme
EDF's~\cite{Inakura11}. In the case of $^{48}$Ca, the experimental
neutron skin thickness ($0.14-0.20$~fm) has been determined from
the $E1$ strength distribution which is extracted from proton
inelastic scattering~\cite{Birkhan16}. HF-BCS analysis gives the
neutron skin $\Delta R_{np}$ of $^{48}$Ca to be 0.16~fm and
0.14~fm with the SLy5 and SLy5+T EDF's, respectively. The
theoretical ``model-averaged'' estimate for $\Delta R_{np}$ is
$0.176{\pm}0.018$~fm~\cite{Piekarewicz12}. In addition, the {\it
ab initio} calculations for the neutron skin in $^{48}$Ca is
$0.12{\leq}\Delta R_{np}{\leq}0.15$~fm~\cite{Hagen16}.

For the interaction in the p-p channel, we use a zero-range volume
force, i.e., $\eta=0$ in Eq.~(\ref{pair}). The strength $V_{0}$ is
taken equal to $-270$~MeV{}fm$^{3}$. This value of the pairing
strength is fitted to reproduce the experimental neutron pairing
energies of $^{50,52,54}$Ca obtained from binding energies of
neighboring isotopes. This choice of the pairing interaction has
also been used for a satisfactory description of the experimental
data of $^{70,72,74,76}$Ni~\cite{Severyukhin14}, $^{90,92}$Zr and
$^{92,94}$Mo~\cite{Severyukhin12}. Thus, hereafter we use the
Skyrme interaction SLy5 with and without tensor components in the
particle-hole channel together with the volume zero-range force
acting in the particle-particle channel.

\begin{figure}[t!]
\includegraphics[width=1.0\columnwidth]{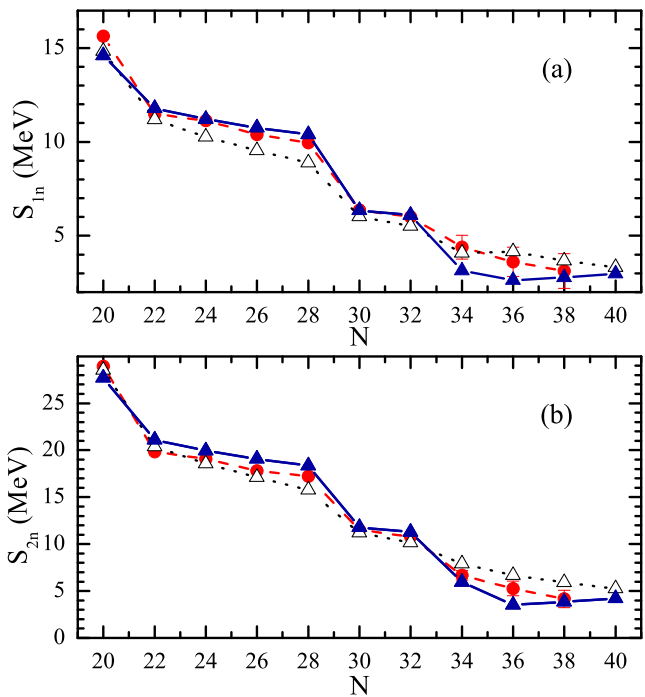}
\caption{(Color online) The one- (top panel) and two- (lower
panel) neutron separation energies for Ca isotopes as a function
of neutron number, calculated with the SLy5 EDF (open triangles)
and with SLy5{+}T (filled triangles). The experimental and
extrapolated energies (filled circles) are from the AME2012 atomic
mass evaluation~\cite{Wang12}.} \label{CaS1nS2n}
\end{figure}

Valuable information about nuclear properties can be obtained from
studies of the one-neutron ($S_{1n}$) and two-neutron ($S_{2n}$)
separation energies. In addition, the separation energies are very
important for the nuclear shell structure. Therefore, it is
interesting to further investigate the evolution of one- and
two-neutron separation energies with both Skyrme interactions. The
neutron separation energies are defined as
\begin{equation}
S_{xn}=B(Z,N)-B(Z,N-X),
\end{equation}
The calculated $S_{1n}$ and $S_{2n}$ values in the even Ca
isotopes are compared with the experimental data~\cite{Wang12} (or
values deduced from systematic trends) in Fig.~\ref{CaS1nS2n}.
Despite the overestimation of the Ca binding energies, the
effective interactions SLy5 and SLy5{+}T give a reasonable
reproduction of the experimental trends in stable nuclei. It
should be noted that the binding energies of the odd Ca isotopes
are calculated with the blocking effect for unpaired
nucleons~\cite{Soloviev61,Dobaczewski09}. For $^{39}$Ca, the
neutron quasiparticle blocking is based on filling the
$\nu1d\frac{3}{2}$ subshell while the $\nu1f\frac{7}{2}$ subshell
should be blocked for $^{41,43,45,47}$Ca. The neutron
$\nu2p\frac{3}{2}$, $\nu2p\frac{1}{2}$ and $\nu1f\frac{5}{2}$
subshells are chosen to be blocked in the cases of the
$^{49,51}$Ca, $^{53}$Ca and $^{55,57,59}$Ca isotopes,
respectively. The existing experimental data show a different
A-behavior, namely, the factor 5 reduction of $S_{1n}$ values and
the 7-time reduction of $S_{2n}$ values from $^{40}$Ca to
$^{58}$Ca. In general, both the SLy5 and SLy5{+}T interactions
give an excellent description of $S_{1n}$ and $S_{2n}$ for
$^{50,52}$Ca isotopes. Most notably, the major shell closure at
the magic neutron number $N=20$ is too pronounced. Also, sharp
decreases in separation energies are seen at the magic neutron
number $N=28$. Significant differences are developed starting from
$^{52}$Ca. In the case of the SLy5 interaction our calculations
predicted a monotonic decrease of $S_{2n}$ when going from
$^{52}$Ca to $^{60}$Ca. On the other hand, the presence of tensor
components brings a drop in the theoretical $S_{2n}$ in
$^{52,54}$Ca~\cite{Grasso14}. This corresponds to the hypothetical
shell closures at $N=32,34$. This suggests that these nuclei are
magic for the used interaction (see in Fig.~\ref{CaS1nS2n}(b)),
thus matching predictions from shell model calculations including
three-body forces~\cite{Holt12,Hagen12}. This view is supported by
precision mass measurements~\cite{Wienholtz13,Steppenbeck13}.
Thus, this jump is a shell effect, and the results indicate that
the various forces lead to different detailed shell structures.

In order to construct the wave functions~(\ref{wf2ph}) of the
$1^{-}$~states, in the present study we take into account all
two-phonon terms that are constructed from the phonons with
multipolarities~$\lambda{\leq}5$~\cite{Arsenyev12,Arsenyev15,Arsenyev16}.
As an example the energies and reduced transition probabilities of
the first $2^{+}$, $3^{-}$, $4^{+}$ and $5^{-}$ phonons for
$^{46,48,50}$Ca are presented in Table~\ref{tab1}. The QRPA
results obtained with the SLy5 EDF are compared with the
experimental data~\cite{Wu00,Kibedi02,Burrows06,Montanari12}. As
one can see, the overall agreement of the energies and
$B(E\lambda)$ values with the data looks reasonable. All dipole
excitations with energies below 35~MeV and 15~most collective
phonons of the other multipolarities are included in the wave
function~(\ref{wf2ph}). We have checked that extending the
configuration space plays a minor role in our calculations.
\begin{table}
\caption{Energies and $B(E\lambda)$ values for up-transitions to
the $\lambda_{1}^{\pi}$ states in $^{46,48,50}$Ca. Experimental
data are taken from
Refs.~\cite{Wu00,Kibedi02,Burrows06,Montanari12}.} \label{tab1}
\begin{ruledtabular}
\begin{tabular}{cccccccc}
 &$\lambda^{\pi}_{1}$&\multicolumn{2}{c}{Energy}&\multicolumn{2}{c}{$B(E\lambda;0^{+}_{gs}\rightarrow\lambda^{\pi}_{1})$}\\
 &         &\multicolumn{2}{c}{(MeV)} &\multicolumn{2}{c}{(e$^2$b$^{\lambda}$)} \\
 &         & Expt.  &  Theory & Expt.        &  Theory         \\
\noalign{\smallskip}\hline\noalign{\smallskip}$^{46}$Ca
 &$2^{+}_{1}$&1.346 & 2.05    & $0.0127{\pm}0.0023$& 0.0070 \\
\noalign{\smallskip}
 &$3^{-}_{1}$&3.614 & 4.57    & $0.006{\pm}0.003$& 0.0049 \\
\noalign{\smallskip}
 &$4^{+}_{1}$&2.575 & 2.30    &                & 0.00035 \\
\noalign{\smallskip}
 &$5^{-}_{1}$&4.184 & 4.67    &                & 0.00027 \\
\noalign{\smallskip}\hline\noalign{\smallskip}$^{48}$Ca
 &$2^{+}_{1}$&3.832 & 3.19    & $0.00968{\pm}0.00105$& 0.0065 \\
\noalign{\smallskip}
 &$3^{-}_{1}$&4.507 & 4.47    & $0.0083{\pm}0.0020$& 0.0038 \\
\noalign{\smallskip}
 &$4^{+}_{1}$&4.503 & 3.51    &                & 0.00035 \\
\noalign{\smallskip}
 &$5^{-}_{1}$&5.729 & 4.52    &                & 0.00026 \\
\noalign{\smallskip}\hline\noalign{\smallskip}$^{50}$Ca
 &$2^{+}_{1}$&1.027 & 1.50    & $0.00375{\pm}0.00010$& 0.0018 \\
\noalign{\smallskip}
 &$3^{-}_{1}$&3.997 & 4.36    &                & 0.0045 \\
\noalign{\smallskip}
 &$4^{+}_{1}$&4.515 & 3.75    &                & 0.00051 \\
\noalign{\smallskip}
 &$5^{-}_{1}$&5.110 & 4.45    &                & 0.00029 \\
\end{tabular}
\end{ruledtabular}
\end{table}

\begin{figure}[t!]
\includegraphics[width=1.0\columnwidth]{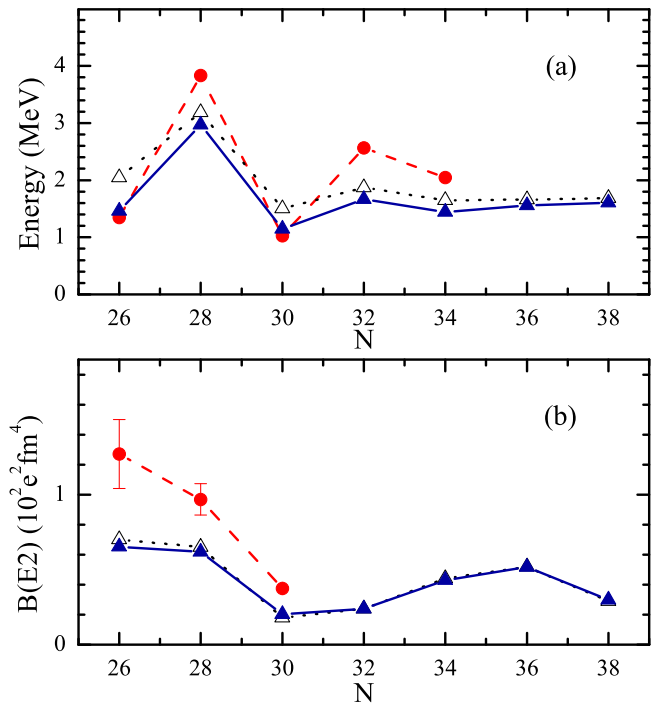}
\caption{(Color online) Energies (top panel) and $B(E2)$ (lower
panel) values for up-transitions to the $2^{+}_{1}$ states in the
neutron-rich Ca isotopes. Results of the calculations within the
QRPA (open triangles) with the SLy5 EDF and the QRPA plus PPC
(filled triangles) are shown. Experimental data (filled circles)
are taken from Refs.~\cite{Gade06,Steppenbeck13,Montanari12}.}
\label{CaBE2}
\end{figure}

It is interesting to examine the energies and transition
probabilities of the first collective phonons which leads to the
minimal two-phonon energy and the maximal matrix elements for
coupling of the one- and two-phonon configurations~(\ref{2phU}).
The calculated $2^{+}_{1}$ energies and transition probabilities
in the neutron-rich Ca isotopes are compared with existing
experimental data~\cite{Gade06,Steppenbeck13,Montanari12} in
Fig.~\ref{CaBE2}. The first $2^{+}$ states of the even-even
$^{46-58}$Ca isotopes exhibit pure neutron two-quasiparticle~(2QP)
excitations ($>72{\%}$). It is worth mentioning that a similar
observation has been found in Ref.~\cite{Terasaki06}, where the
SkM$^{\ast}$ interaction was used. There is a marked increase of
the $2^{+}_{1}$ energy of $^{48}$Ca in comparison with those in
$^{46}$Ca and $^{50}$Ca. It corresponds to a standard evolution of
the $2^{+}_{1}$ energy near closed shells. It should be noted that
including the tensor components changes the contributions of the
main configurations only slightly, but the general structure of
the $2^{+}_{1}$ state remains the same. In the following, we
analyze the PPC effects in the case of the first $2^{+}$
excitations. In these calculations all two-phonon configurations
below 20~MeV are used. In $^{46-58}$Ca isotopes, the crucial
contribution in the wave function structure of the first
$2^{+}$~state comes from the $[2^{+}_{1}]_{QRPA}$ ($>89{\%}$)
configuration, i.e., $R_{i{=}1}^{2}(\lambda{=}2\nu{=}1){>}0.89$.
Using the set of linear equations~(\ref{2pheq1}) and
(\ref{2pheq2}), one can get $E_{1}{\leq}\omega_{21}$. Since the
transition matrix elements for direct excitation of two-phonon
components from the ground state are about two orders of magnitude
smaller as compared to the excitation of one-phonon components, it
follows that $B(E2)_{PPC}{\leq}B(E2)_{QRPA}$ in the case of the
first $2^{+}$~state. The decrease of the $2^+_1$ energies and the
$B(E2)$ values is shown in Fig.~\ref{CaBE2}. It is worth to
mention that the first discussion of the PPC effect for the
$2^{+}_1$ properties based on QRPA calculations with Skyrme forces
has been done in Ref.~\cite{Severyukhin04}. Fig.~\ref{CaBE2}
demonstrates that the calculated energies and $B(E2)$ values of
the first $2^{+}$~excited states deviate from the experimental
data. In the case of $^{46}$Ca, the $B(E2)$ value is about
1.5~times less than the data. This probably points to a particular
problem due to the EDF used here rather than to a deficiency of
our model space.
%

%
%
\section{Results and discussion}

\begin{figure}[t!]
\includegraphics[width=1.0\columnwidth]{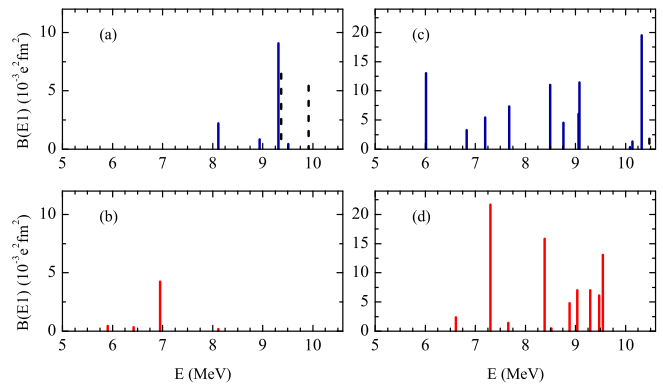}
\caption{(Color online) Low-energy $E1$ strength distributions of
$^{40}$Ca (resp. $^{48}$Ca) are shown in the left (resp. right)
panels. The dashed and solid lines correspond to the SLy5
calculations within the RPA and taking into account the PPC
effects, respectively. Panels (b) and (d): experimental data are
from Ref.~\cite{Derya14}.} \label{Ca40Ca48BE1}
\end{figure}
\begin{figure}[t!]
\includegraphics[width=1.0\columnwidth]{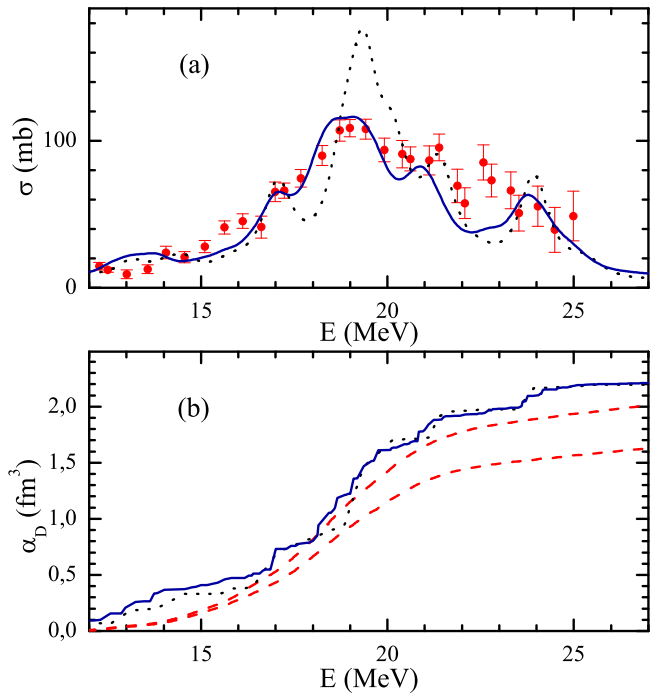}
\caption{(Color online) (a) The estimated photo-absorption cross
section for $^{48}$Ca (filled circles) are taken from
Ref.~\cite{Keefe87}. The dotted and solid lines correspond to the
calculations within the RPA with the SLy5 EDF and taking into
account the PPC, respectively. (b) Running sum of the electric
dipole polarizability for $^{48}$Ca calculated within the RPA with
the SLy5 EDF (dotted line) and the RPA plus PPC (solid line) in
comparison to experimental determination of $\alpha_{D}$ (the two
dashed lines indicate upper and lower limits)~\cite{Birkhan16}.}
\label{Ca48GDR}
\end{figure}

As a first step in the present analysis, we examine the PPC
effects on the $E1$ strength distributions for doubly-magic
$^{40,48}$Ca isotopes. A comparison of such calculations with
recent experimental data~\cite{Derya14} demonstrates that the RPA
approach cannot reproduce correctly the low-energy $E1$ strength
distributions, see Fig.~\ref{Ca40Ca48BE1}. Let us discuss the
properties of the lowest dipole state. For $^{48}$Ca this state is
mainly characterized by the two-phonon component
$[2^{+}_{1}\otimes3^{-}_{1}]$ arising from the coupling between
the first quadrupole and octupole phonons. We find a nice
agreement with the data, where the candidate for the two-phonon
$1^{-}$ state is expected at an energy of 7.298~MeV with
$B(E1;0^{+}_{gs}\rightarrow{1^{-}_{1}})=18.6{\pm}1.8{\times}10^{-3}$e$^2$fm$^{2}$
\cite{Hartmann00}. For $^{40}$Ca the PPC calculation predicts the
first $1^{-}$ state significantly higher than the experimental
two-phonon candidate~\cite{Derya16} (see Fig.~\ref{Ca40Ca48BE1}).
It is worth pointing out that our results for $^{40,48}$Ca are in
good agreement with the RQTBA calculations taking into account the
effects of coupling between quasiparticles and
phonons~\cite{Egorova16}. In addition, we discuss the GDR energy
region. For $^{48}$Ca, the photo-absorption process is well
studied experimentally. The photo-absorption cross section up to
27~MeV is displayed in Fig.~\ref{Ca48GDR}(a). The cross section is
computed by using a Lorentzian smearing with an averaging
parameter $\Delta=1.0$~MeV. The PPC effects yield a noticeable
redistribution of the GDR strength in comparison with the RPA
results. It is worth mentioning that the coupling increases the
GDR width from 6.9~MeV to 7.3~MeV in the energy region
$10-26$~MeV. Also, the PPC induces a 300~keV downward shift of the
GDR energy (19.3~MeV for the RPA). The experimental GDR width and
energy are 6.98~MeV and 19.5~MeV~\cite{Keefe87}, respectively. The
calculated characteristics of the GDR are in agreement with the
observed values. The general shapes of the GDR obtained in the PPC
are rather close to those observed in experiment. This
demonstrates the improvement of the PPC description in comparison
with RPA. We conclude that the main mechanisms of the GDR
formation in $^{48}$Ca can be taken into account correctly and
consistently in the PPC approach.

As proposed in Ref.~\cite{Lipparini89}, we can estimate the $M1$
contribution to the photo-absorption process calculated in the
case of the SLy5 EDF. For $^{48}$Ca, we find that the $M1$
contribution  plays a minor role ($<1{\%}$) for the integrated
cross section. It was shown in the experimental
papers~\cite{Hartmann04,Isaak11} that the contribution of $1^{+}$
components to the total dipole strength below 10~MeV is
negligible. For $^{44}$Ca, the contribution of $M1$ in the region
of $3-10$~MeV to the total dipole strength is less than
3{\%}~\cite{Isaak11}. In the GDR energy region, the $M1$
contribution was zero within error bars and was replaced by a
phenomenological background in the case of
$^{208}$Pb~\cite{Tamii11}. Thus, the magnetic counterparts have
been omitted in our analysis.

In order to perform further investigations on the $^{48}$Ca
nucleus we have extracted the electric dipole
polarizability~\cite{Migdal44,Bohigas81,Satula06}, which
represents a handle to constrain the equation of state of neutron
matter and the physics of neutron stars~\cite{Brown00,Horowitz01}.
The electric dipole polarizability $\alpha_{D}$ is written as
\begin{equation}
\alpha_{D}=\frac{8\pi}{9}\sum\limits_{\mu_{\alpha}\mu}\sum\limits_{i_{\alpha}}
E^{-1}_{1^{-}_{i_{\alpha}}}
\left[\langle1^{-}_{i_{\alpha}\mu_{\alpha}}|\hat{M}^{1\mu}|0^{+}_{gs}\rangle\right]^{2},
\end{equation}
where
\begin{equation}
\hat{M}^{1\mu}=-\frac{Z}{A}e\sum\limits_{k=1}^{N}r_{k}Y_{1\mu}(\hat{r}_k)+
\frac{N}{A}e\sum\limits_{k=1}^{Z}r_{k}Y_{1\mu}(\hat{r}_k).
\label{me1}
\end{equation}
Here, $N$, $Z$, and $A$ are the neutron, proton, and mass numbers,
respectively; $r_{k}$ indicates the radial coordinate for neutrons
(protons); and $Y_{1\mu}(\hat{r}_k)$ is the corresponding
spherical harmonic. The definition of the dipole operator
eliminates contaminations of the physical response due to the
spurious excitation of the center of mass. In
Ref.~\cite{Arsenyev10} it has been shown that eliminating the
spurious state by means of effective charges or the alternative
ways lead to very similar results.

Running sums of $\alpha_{D}$ values for $^{48}$Ca in the energy
region below 27~MeV are given in Fig.~\ref{Ca48GDR}(b). It is
shown that the PPC does not affect the description of the electric
dipole polarizability. The results differ insignificantly.
Moreover, we have checked that inclusion of the tensor components
does not change the value of $\alpha_{D}$ obtained by integrating
the $E1$~strength up to 60~MeV: $\alpha_{D}=2.28$~fm$^{3}$ in the
case of the SLy5 EDF and $\alpha_{D}=2.20$~fm$^{3}$ in the
SLy5{+}T. Both effective interactions reproduce the experimental
data $\alpha_{D}=2.07{\pm}0.22$~fm$^{3}$~\cite{Birkhan16} and they
are in good agreement with the ``model-averaged'' value of
$2.306{\pm}0.089$~fm$^{3}$, which is predicted in
Ref.~\cite{Piekarewicz12}. Although the GDR strength dominates,
contributions to $\alpha_{D}$ value at lower and higher excitation
energies must be taken into account.

\begin{figure}[t!]
\includegraphics[width=1.0\columnwidth]{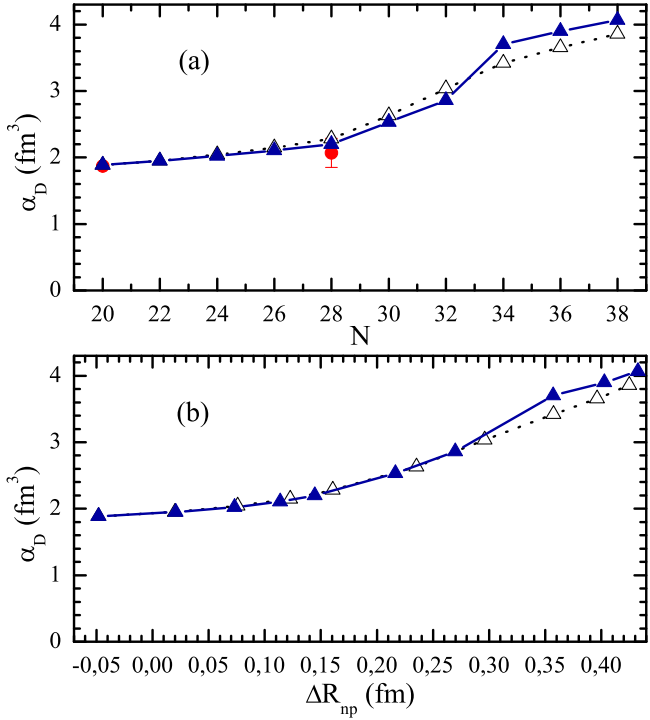}
\caption{(Color online) (a) The electric dipole polarizability
$\alpha_{D}$ as a function of neutron number, calculated with the
SLy5 EDF (open triangles) and with SLy5{+}T (filled triangles).
Experimental values (filled circles) of $\alpha_{D}$ are taken
from~\cite{Birkhan16}. (b) The same as (a) but as a function of
the neutron skin thickness $\Delta R_{np}$.} \label{CaalphaD}
\end{figure}
To complete the discussion we consider $\alpha_{D}$ as a function
of the neutron number for Ca isotopes, see Fig.~\ref{CaalphaD}.
The result of the SLy5 calculation with the PPC predicts a
monotonic increase of the $\alpha_{D}$ value with neutron number,
and only a small kink in the calculated excitation energies is
found at the $N=28$ shell closure. The calculated polarizabilities
$\alpha_{D}$ of $^{40}$Ca and $^{48}$Ca are in excellent agreement
with the experimental data~\cite{Birkhan16}. As shown in
Fig.~\ref{CaalphaD}(a), the SLy5 and SLy5{+}T EDF's produce
qualitatively the same results. We find that the correlation
between the value of $\alpha_{D}$ and neutron skin $\Delta R_{np}$
is discerned, see Fig.~\ref{CaalphaD}(b). With the increase of the
neutron skin one observes a smooth increase of $\alpha_{D}$. Thus,
the $\alpha_{D}$ value and the neutron skin $\Delta R_{np}$ are
correlated as predicted in Ref.~\cite{Satula06}. Besides this, the
$\alpha_{D}$ value can be measured in finite nuclei and, as a
result, the $\Delta R_{np}$ value can be extracted.

\begin{figure}[t!]
\includegraphics[width=1.0\columnwidth]{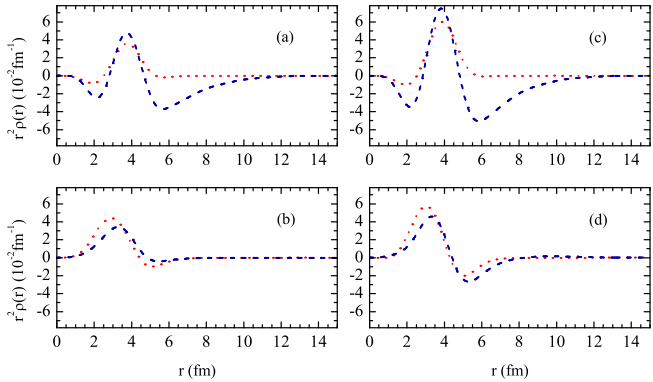}
\caption{(Color online) Transition proton (dotted line) and
neutron (dashed line) densities to selected QRPA $1^-$ states of
$^{50}$Ca (resp. $^{56}$Ca) in the left (resp. right) panels.
Panels (a) and (b) correspond to the transition densities for the
states at 9.1~MeV and 10.3~MeV in $^{50}$Ca, while panels (c) and
(d) show the transition densities for the QRPA $1^-$ states at
8.7~MeV and 10.7~MeV in $^{56}$Ca, respectively. All transition
densities are multiplied by $r^{2}$.} \label{Ca50Ca56rho}
\end{figure}
For $^{48-58}$Ca, we recognize the PDR energy below 10~MeV. The
difference of the structure of the $1^-$ states below and above
10~MeV is illustrated with the QRPA transition densities in the
cases of $^{50}$Ca and $^{56}$Ca, see Fig.~\ref{Ca50Ca56rho}. For
the $1^-$ states below 10~MeV, the proton and neutron densities in
the nuclear interior region are in phase. However one can see the
neutron dominance when $r>6$~fm. For the $1^-$ states about
10~MeV, the corresponding transition proton and neutron densities
are very similar to each other (see Fig.~\ref{Ca50Ca56rho}). These
states can be identified as an isoscalar mode in the low-energy
$E1$ strength. This is not the PDR where the neutron skin
oscillates against the inner core. The next fairly collective
$1^-$ state at 12.2~MeV in $^{50}$Ca and 12.1~MeV in $^{56}$Ca has
neither the isoscalar character nor the isovector character.
Increasing further the excitation energy we observe the low-energy
GDR tail. Note that the PDR energy region for the neutron-rich Ca
isotopes has been found in the framework of other theoretical
approaches~\cite{Inakura11,Egorova16,Vretenar01}.

\begin{figure}[t!]
\includegraphics[width=1.0\columnwidth]{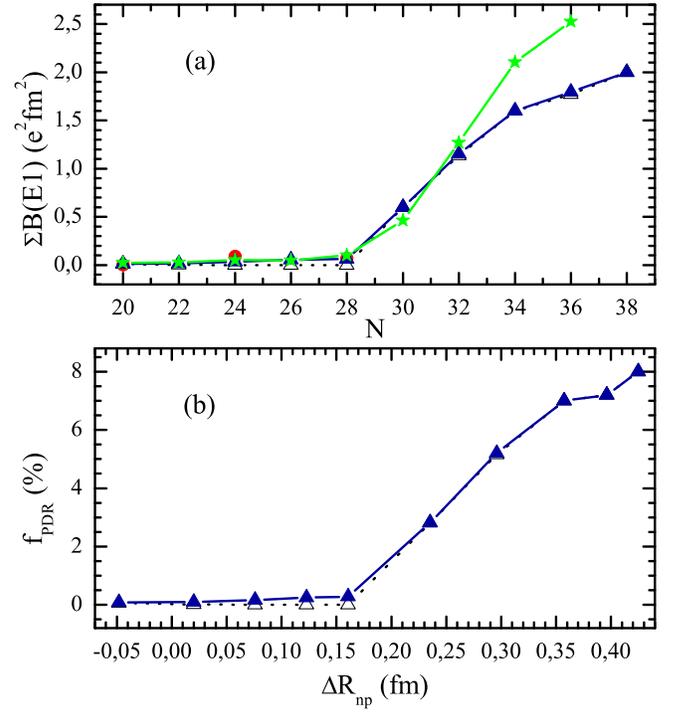}
\caption{(Color online) (a) Summed dipole strength below 10~MeV
calculated in the QRPA (open triangles) with the SLy5 EDF, PPC
(filled triangles) and with RQTBA (filled stars)~\cite{Egorova16}.
Experimental data (filled circles) are taken from
Refs.~\cite{Hartmann04}. (b) Ratio of the PDR energy-weighted
strength to the TRK sum rule for various Ca isotopes as a function
of the neutron skin thickness~$\Delta R_{np}$. The open triangles
indicate discrete QRPA results, while the filled triangles are PPC
results. } \label{CaFpdr}
\end{figure}
Let us now discuss the low-energy $E1$ strength. The collectivity
of the PDR can be studied by plotting the evolution of its summed
strength $\sum{B(E1)}$ with respect to the mass number. The
calculated PDR is integrated up to 10~MeV. As shown in the upper
panel of Fig.~\ref{CaFpdr}, the behavior of the PDR summed
strength can be divided into two categories: light and heavy Ca
isotopes. In particular, we find that there is a sharp increase
after the doubly magic isotope $^{48}$Ca in the QRPA with the SLy5
EDF. In light Ca isotopes a completely different behavior is
observed in Ref.~\cite{Hartmann04}: the measured summed strength
in $^{44}$Ca is 18~times larger than in $^{40}$Ca and 1.3~times in
the case of $^{48}$Ca. From Fig.~\ref{CaFpdr}(a), one can see that
the QRPA calculations fail to reproduce the experimental data.
This result is in agreement with the
nonrelativistic~\cite{Terasaki06} and relativistic
QRPA~\cite{Vretenar01} calculations. Thus, the correlation between
the PDR integrated strength and the neutron excess of Ca isotopes
is nontrivial and it is necessary to go beyond QRPA to explain the
properties of the PDR.

\begin{figure}[t!]
\includegraphics[width=1.0\columnwidth]{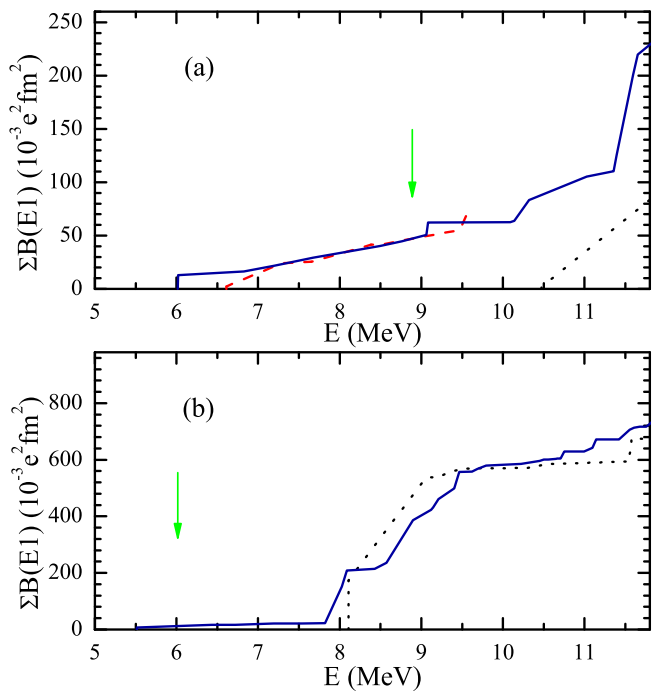}
\caption{(Color online) Running sums of the low-lying dipole
strengths in $^{48}$Ca (top panel) and $^{50}$Ca (lower panel).
The dotted and solid lines correspond to the calculations within
the QRPA with the SLy5 EDF and taking into account the PPC,
respectively. Experimental data (dashed line) are taken from
Ref.~\cite{Hartmann04}. The calculated $S_{1n}$ energy is
indicated by the solid arrow.} \label{Ca48Ca50SUM}
\end{figure}
Let us now discuss the strong increase of the summed $E1$~strength
below 10~MeV ($\sum{B(E1)}$), with increasing neutron number from
$^{48}$Ca till $^{58}$Ca. In Fig.~\ref{Ca48Ca50SUM}(a) the
calculated running sum for $^{48}$Ca is plotted as a function of
the excitation energy. In the same plot the calculated $S_{1n}$
values are shown. In the case of the RPA, there is no $1^{-}$
state below 10~MeV, see Fig.~\ref{Ca40Ca48BE1}. The RPA
calculations predict the first dipole state around 10.5~MeV. In
contrast to the RPA case, the PPC results in the formation of
low-lying $1^{-}$~states in this energy region. The dominant
contribution in the wave function of the $1^{-}$ states comes from
the two-phonon configurations ($>60{\%}$). These states originate
from the fragmentation of the RPA states above 10~MeV. As one can
see in Fig.~\ref{Ca48Ca50SUM}(a), the calculated running sum of
the $\sum{B(E1)}$ value is close to the experimental $\sum{B(E1)}$
value. The PPC calculations give a total dipole strength of
0.063~e$^2$fm$^2$. The experimental value of $\sum{B(E1)}$ is
$0.0687{\pm}0.0075$~e$^2$fm$^2$ in the same
interval~\cite{Hartmann04}. The PPC effects produce a sizable
impact on the low-energy $E1$ strength of $^{48}$Ca. It is
remarkable that the contributions of the low-lying $1^{-}$~states
to the value of $\alpha_{D}$ is small (0.033~fm$^3$), three times
as much as the value deduced experimentally~\cite{Birkhan16}. It
is shown that the $\alpha_{D}$ value is more sensitive to the fine
structure of the $E1$~strength distribution. The PPC calculations
reproduce the observed trend in light Ca isotopes, although the
theoretical value of $\sum{B(E1)}$ for $^{44}$Ca underestimates
the experimental value by a factor of 2. It is worth mentioning
that the relativistic quasiparticle time blocking
approximation~(RQTBA)~\cite{Egorova16}, the value of $\sum{B(E1)}$
is also substantially less than the experimental one. As shown in
Fig.~\ref{CaFpdr}(a), the two models produce qualitatively the
same results.

\begin{figure}[t!]
\includegraphics[width=1.0\columnwidth]{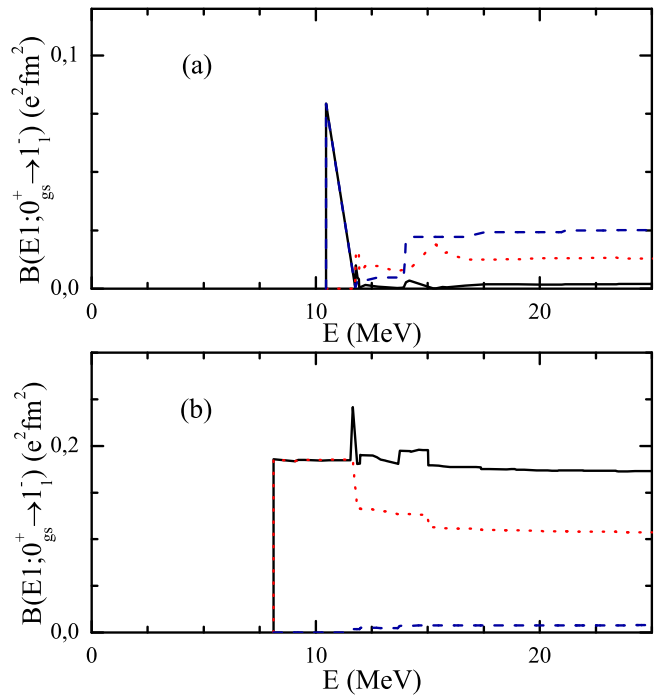}
\caption{(Color online) Running sums (solid lines) of the
$B(E1;0^{+}_{gs}\rightarrow 1^{-}_{1})$ as a function of the
two-quasiparticle energy included in the QRPA calculations for
$^{48}$Ca (top panel) and $^{50}$Ca (lower panel). Dashed and
dotted lines correspond to the results for the proton and the
neutron components of the dipole operator~(\ref{me1}),
respectively.} \label{Ca48Ca50BE1SUM}
\end{figure}
Moving from $^{48}$Ca to $^{50}$Ca, the QRPA calculations predict
a jump of the $\sum{B(E1)}$ value. $N=30$ corresponds to the
occupation of $\nu2p\frac{3}{2}$ subshell, resulting in the two
rather pronounced states below 10~MeV being pure neutron 2QP
excitations: $99{\%}\{3s\frac{1}{2}2p\frac{3}{2}\}_{\nu}$ and
$98{\%}\{2d\frac{5}{2}2p\frac{3}{2}\}_{\nu}$. The summed $E1$
strength of these states is 0.54~e$^2$fm$^2$. As can be seen from
Fig.~\ref{Ca48Ca50SUM}(b), two states determine the value of
$\sum{B(E1)}$ calculated below 10~MeV. There is no contribution
from the 2QP proton excitations. This structure is very different
from that of the first $1^{-}$ state in $^{48}$Ca, where the
leading proton 2QP configuration
$\{2p\frac{3}{2}1d\frac{3}{2}\}_{\pi}$ gives a contribution of
96{\%}. For $^{48}$Ca, the closure of the neutron subshell
$\nu1f\frac{7}{2}$ leads to the vanishing of the neutron pairing.

Fig.~\ref{Ca48Ca50BE1SUM} shows the running sums for the QRPA
value of $B(E1;0^{+}_{gs}\rightarrow 1^{-}_{1})$ as a function of
the proton and neutron 2QP energies. In the case of $^{48}$Ca, the
$B(E1;0^{+}_{gs}\rightarrow 1^{-}_{1})$ is exhausted by the proton
2QP configurations, while the largest part of the $B(E1)$ value
for $^{50}$Ca is generated by neutron excitations in the
low-energy region. For $^{50}$Ca, the proton and neutron
components contribute coherently. The main difference between
these isotopes is that the neutron 2QP configurations contribute
more than proton ones. This competition is mainly responsible for
the $\sum{B(E1)}$ increase.

In contrast to the case of $^{48}$Ca, the PPC effects on the
low-energy dipole spectrum of $^{50}$Ca is weak (see
Fig.~\ref{Ca48Ca50SUM}(b)). Thus, the $\sum{B(E1)}$ values for
$^{50}$Ca results predominantly from the QRPA distribution of
$E1$~strength. As can be seen in Fig.~\ref{CaFpdr}(a) a similar
result is observed in the case of $^{52,54,56,58}$Ca isotopes. The
separation energies decrease much faster than the value of
$\sum{B(E1)}$. This means, of course, that the observation of the
PDR in $(\gamma,\gamma^{\prime})$ experiments will be strongly
hindered.

Let us now examine the correlation between the PDR properties and
the neutron skin $\Delta R_{\rm np}$. To quantify the low-energy
$E1$ strength in a systematic analysis, we use the PDR fraction,
\begin{eqnarray}
f_{PDR}=\nonumber\\
\frac{\sum\limits_{i_{\alpha}}^{E_{1^{-}_{i_{\alpha}}}{\leq}10~{MeV}}
E_{1^{-}_{i_{\alpha}}}\sum\limits_{\mu_{\alpha}\mu}
\left[\langle1^{-}_{i_{\alpha}\mu_{\alpha}}|\hat{M}^{1\mu}|0^{+}_{gs}\rangle\right]^{2}}
{14.8{\cdot}NZ/A~e^{2}{\rm{fm^{2}}{\cdot}MeV}}.
\end{eqnarray}
One of the basic ingredients for the fitting protocol of the SLy5
EDF is the enhancement factor of the TRK sum rule
$\kappa=0.25$~\cite{Chabanat98}. Therefore, the total dipole
strength exhausts 125{\%} of the TRK sum rule, i.e.,
$18.5{\cdot}NZ/A~e^{2}{\rm{fm^{2}}{\cdot}MeV}$. In
Fig.~\ref{CaFpdr}(b) the $\Delta R_{\rm np}$ dependence of the
$f_{PDR}$ for Ca isotopes is shown. The filled triangles indicate
the PPC effects, which can be compared with QRPA results indicated
by the open triangles. The $^{48}$Ca$(\gamma,\gamma^{\prime})$
experiments give $f_{PDR}=0.33{\pm}0.04${\%}~\cite{Hartmann04},
while the calculations with the PPC effects lead to 0.28{\%} (to
be compared with the RQTBA result of 0.55{\%}~\cite{Egorova16}).
The QRPA calculations predict that the visible $f_{PDR}$ kink can
be identified at the magic neutron number $N=28$. The PDR
fractions suddenly increase from $N=28$ to 30 and continue to
increase until $N=34$ where the $\nu2p\frac{1}{2}$ subshell is
filled. Beyond $N=34$, the neutrons start filling the
$\nu1f\frac{5}{2}$ subshell, thus reducing the slope of $f_{PDR}$.
The correlation between the $f_{PDR}$ and $\Delta R_{\rm np}$ turn
out to be rather complex, depending on the neutron number.

%
%
\section{Conclusions}

The electric dipole polarizability is a particularly important
observable, as it can be measured in finite nuclei and it provides
important information on the neutron skin thickness that can be
extracted. In this study, Skyrme QRPA calculations including the
phonon-phonon coupling have been performed for the $E1$ response
in neutron-rich Ca isotopes, some of which should become
experimentally accessible in the near future. The FRSA enables one
to perform the calculations in large configuration spaces. The
SLy5 EDF and its modification including the tensor components
SLy5{+}T reproduce the data for the neutron skin thickness and
neutron separation energies. Among our initial motivation there
was the estimation of dipole polarizability for $^{50,52}$Ca in
comparison to the $N=28$ isotope $^{48}$Ca. Our results describe
the experimental data of $^{40,48}$Ca and they give a sizable
increase due to the neutron shell closure effect. It is shown that
the phonon-phonon coupling has small influence on the dipole
polarizability.

For $^{48}$Ca, the PPC effect have a damping and smoothing action
which yields a GDR cross section close to the experimental one in
shape and magnitude. We find the impact of the shell closure
$N=28$ on the summed $E1$~strength below 10~MeV. There is also the
64{\%} decrease of one-neutron separation energy from $^{48}$Ca to
$^{50}$Ca. This is due to the pairing effect on neutron
$\nu2p\frac{3}{2}$ subshell for $^{50}$Ca. The dipole response for
$^{52-58}$Ca is characterized by the fragmentation of the strength
distribution and its spreading into the low-energy region.

The model can be extended by enlarging the variational space for
the $1^{-}$ states with the inclusion of the three-phonon
configurations. The computational developments that would allow us
to conclude on this point are underway.

%
%
\section*{Acknowledgments}

We are thankful to I. A. Egorova, N. Pietralla, M. Scheck and Ch.
Stoyanov for many fruitful and stimulating discussions concerning
various aspect of this work. The authors are very thankful to E.
O. Sushenok for help in calculations of the neutron separation
energies. N.N.A., A.P.S., and V.V.V. thank the hospitality of
IPN-Orsay where the part of this work was done. This work is
partly supported by the CNRS-RFBR agreement No.~16-52-150003, the
IN2P3-JINR agreement, and the RFBR grant No.~16-02-00228.

%
%

%
%

\begin{thebibliography}{99}
\bibitem{Migdal44}
A. B. Migdal, J. Phys. Acad. Sci. USSR {\bf 8}, 331 (1944); J.
Exp. Theor. Phys. USSR {\bf 15}, 81 (1945).

\bibitem{Berman75}
B. L. Berman and  S. C. Fultz, Rev. Mod. Phys. {\bf 47}, 713
(1975).

\bibitem{Dietrich88}
S. S. Dietrich and B. L. Berman, At. Data Nucl. Data Tables {\bf
38}, 199 (1988).

\bibitem{Savran13}
D. Savran, T. Aumann, and A. Zilges, Prog. Part. Nucl. Phys. {\bf
70}, 210 (2013).

\bibitem{Paar07}
N. Paar, D. Vretenar, E. Khan, and G. Col\`{o}, Rep. Prog. Phys.
{\bf 70}, 691 (2007).

\bibitem{Inakura11}
T. Inakura, T. Nakatsukasa, and K. Yabana, Phys. Rev. C {\bf 84},
021302(R) (2011).

\bibitem{Ebata14}
S. Ebata, T. Nakatsukasa, and T. Inakura, Phys. Rev. C {\bf 90},
024303 (2014); {\bf 92}, 049902(E) (2015).

\bibitem{Arnould07}
M. Arnould, S. Goriely, and  K. Takahashi, Phys. Rep. {\bf 450},
97 (2007).

\bibitem{Brown00}
B. A. Brown, Phys. Rev. Lett. {\bf 85}, 5296 (2000).

\bibitem{Horowitz01}
C. J. Horowitz and J. Piekarewicz, Phys. Rev. Lett. {\bf 86}, 5647
(2001).

\bibitem{Hartmann00}
T. Hartmann, J. Enders, P. Mohr, K. Vogt, S. Volz, and A. Zilges,
Phys. Rev. Lett. {\bf 85}, 274 (2000); {\bf 86}, 4981(E) (2001).

\bibitem{Hartmann04}
T. Hartmann, M. Babilon, S. Kamerdzhiev, E. Litvinova, D. Savran,
S. Volz, and A. Zilges, Phys. Rev. Lett. {\bf 93}, 192501 (2004).

\bibitem{Isaak11}
J. Isaak, D. Savran, M. Fritzsche, D. Galaviz, T. Hartmann, S.
Kamerdzhiev, J. H. Kelley, E. Kwan, N. Pietralla, C. Romig, G.
Rusev, K. Sonnabend, A. P. Tonchev, W. Tornow, and A. Zilges,
Phys. Rev. C {\bf 83}, 034304 (2011).

\bibitem{Derya14}
V. Derya, D. Savran, J. Endres, M. N. Harakeh, H. Hergert, J. H.
Kelley, P. Papakonstantinou, N. Pietralla, V. Yu. Ponomarev, R.
Roth, G. Rusev, A. P. Tonchev, W. Tornow, H. J. W\"{o}rtche, and
A.Zilges, Phys. Lett. B {\bf 730}, 288 (2014).

\bibitem{Soloviev78}
V. G. Soloviev, Ch. Stoyanov, and V. V. Voronov, Nucl. Phys. A
{\bf 304}, 503 (1978).

\bibitem{Terasaki06}
J. Terasaki and J. Engel, Phys. Rev. C {\bf 74}, 044301 (2006).

\bibitem{Egorova16}
I. A. Egorova and E. Litvinova, Phys. Rev. C {\bf 94}, 034322
(2016).

\bibitem{Gade06}
A. Gade, R. V. F. Janssens, D. Bazin, R. Broda, B. A. Brown, C. M.
Campbell, M. P. Carpenter, J. M. Cook, A. N. Deacon, D.-C. Dinca,
B. Fornal, S. J. Freeman, T. Glasmacher, P. G. Hansen, B. P. Kay,
P. F. Mantica, W. F. Mueller, J. R. Terry, J. A. Tostevin, and S.
Zhu, Phys. Rev. C {\bf 74}, 021302(R) (2006).

\bibitem{Wienholtz13}
F. Wienholtz, D. Beck, K. Blaum, Ch. Borgmann, M. Breitenfeldt, R.
B. Cakirli, S. George, F. Herfurth, J. D. Holt, M. Kowalska, S.
Kreim, D. Lunney, V. Manea, J. Men\'{e}ndez, D. Neidherr, M.
Rosenbusch, L. Schweikhard, A. Schwenk, J. Simonis, J. Stanja, R.
N.Wolf, and K. Zuber, Nature {\bf 498}, 346 (2013).

\bibitem{Steppenbeck13}
D. Steppenbeck, S. Takeuchi, N. Aoi, P. Doornenbal, M. Matsushita,
H. Wang, H. Baba, N. Fukuda, S. Go, M. Honma, J. Lee, K. Matsui,
S. Michimasa, T. Motobayashi, D. Nishimura, T. Otsuka, H. Sakurai,
Y. Shiga, P.-A. S\"{o}derstr\"{o}m, T. Sumikama, H. Suzuki, R.
Taniuchi, Y. Utsuno, J. J. Valiente-Dob\'{o}n, and K. Yoneda,
Nature {\bf 502}, 207 (2013).

\bibitem{Tarasov09}
O. B. Tarasov, D. J. Morrissey, A. M. Amthor, T. Baumann, D.
Bazin, A. Gade, T. N. Ginter, M. Hausmann, N. Inabe, T. Kubo, A.
Nettleton, J. Pereira, M. Portillo, B. M. Sherrill, A. Stolz, and
M. Thoennessen, Phys. Rev. Lett. {\bf 102}, 142501 (2009).

\bibitem{Soloviev92}
V. G. Soloviev, {\it Theory of Atomic Nuclei: Quasiparticles and
Phonons} (Institute of Physics, Bristol and Philadelphia, 1992).

\bibitem{LoIudice12}
N. Lo Iudice, V. Yu. Ponomarev, Ch. Stoyanov, A. V. Sushkov, and
V. V. Voronov, J. Phys. G {\bf 39}, 043101 (2012).

\bibitem{Grinberg94}
M. Grinberg and Ch. Stoyanov, Nucl. Phys. A \textbf{573}, 231
(1994).

\bibitem{Ponomarev98}
V.Yu. Ponomarev, Ch. Stoyanov, N. Tsoneva, and M. Grinberg,
Nucl.Phys. A \textbf{635}, 470 (1998).

\bibitem{Tsoneva04}
N. Tsoneva, H. Lenske, and Ch. Stoyanov, Phys. Lett. B
\textbf{586}, 213 (2004).

\bibitem{Tsoneva08}
N. Tsoneva and H. Lenske, Phys. Rev. C \textbf{77}, 024321 (2008).

\bibitem{Tsoneva15}
N. Tsoneva, S. Goriely, H. Lenske, and R. Schwengner, Phys. Rev. C
\textbf{91}, 044318 (2015).

\bibitem{Sarchi04}
D. Sarchi, P. F. Bortignon, and G. Col\`{o}, Phys. Lett. B {\bf
601}, 27 (2004).

\bibitem{Avdeenkov11}
A. Avdeenkov, S. Goriely, S. Kamerdzhiev, and S. Krewald, Phys.
Rev. C {\bf 83}, 064316 (2011).

\bibitem{Arsenyev12}
N. N. Arsenyev, A. P. Severyukhin, V. V. Voronov, and Nguyen Van
Giai, EPJ Web of Conf. {\bf 38}, 17002 (2012).

\bibitem{Repko13}
A. Repko, P.-G. Reinhard, V. O. Nesterenko, and J. Kvasil, Phys.
Rev. C {\bf 87}, 024305 (2013).

\bibitem{Giai98}
Nguyen Van Giai, Ch. Stoyanov, and V. V. Voronov, Phys. Rev. C
{\bf 57}, 1204 (1998).

\bibitem{Severyukhin02}
A. P. Severyukhin, Ch. Stoyanov, V. V. Voronov, and Nguyen Van
Giai, Phys. Rev. C {\bf 66}, 034304 (2002).

\bibitem{Severyukhin04}
A. P. Severyukhin, V. V. Voronov, and Nguyen Van Giai, Eur. Phys.
J. A {\bf 22}, 397 (2004).

\bibitem{Severyukhin08}
A. P. Severyukhin, V. V. Voronov, and Nguyen Van Giai, Phys. Rev.
C {\bf 77}, 024322 (2008).

\bibitem{Suzuki81}
T. Suzuki and H. Sagawa, Prog. Theor. Phys. {\bf 65}, 565 (1981).

\bibitem{Sarriguren99}
P. Sarriguren, E. Moya de Guerra, and A. Escuderos, Nucl. Phys. A
{\bf 658}, 13 (1999).

\bibitem{Nesterenko02}
V. O. Nesterenko, J. Kvasil, and P.-G. Reinhard, Phys. Rev. C {\bf
66}, 044307 (2002).

\bibitem{Severyukhin09}
A. P. Severyukhin, N. N. Arseniev, V. V. Voronov, and Nguyen Van
Giai, Phys. At. Nucl. {\bf 72}, 1149 (2009).

\bibitem{Severyukhin12}
A. P. Severyukhin, N. N. Arsenyev, and N. Pietralla, Phys. Rev. C
{\bf 86}, 024311 (2012).

\bibitem{Arsenyev15}
N. N. Arsenyev, A. P. Severyukhin, V. V. Voronov, and Nguyen Van
Giai, Acta Phys. Pol. B {\bf 46}, 517 (2015).

\bibitem{Arsenyev16}
N. N. Arsenyev, A. P. Severyukhin, V. V. Voronov, and Nguyen Van
Giai, EPJ Web of Conf. \textbf{107}, 05006 (2016).

\bibitem{Ring80}
P. Ring and P. Schuck, {\it The Nuclear Many Body Problem}
(Springer, Berlin, 1980).

\bibitem{Stancu77}
F. Stancu, D. M. Brink, and H. Flocard, Phys. Lett. B {\bf 68},
108 (1977).

\bibitem{Lesinski07}
T. Lesinski, M. Bender, K. Bennaceur, T. Duguet, and J. Meyer,
Phys. Rev. C {\bf 76}, 014312 (2007).

\bibitem{Terasaki05}
J. Terasaki, J. Engel, M. Bender, J. Dobaczewski, W. Nazarewicz,
and M. Stoitsov, Phys. Rev. C {\bf 71}, 034310 (2005).

\bibitem{Severyukhin14}
A. P. Severyukhin, V. V. Voronov, I. N. Borzov, N. N. Arsenyev,
and Nguyen Van Giai, Phys. Rev. C {\bf 90}, 044320 (2014).

\bibitem{Chabanat98}
E. Chabanat, P. Bonche, P. Haensel, J. Meyer, and R. Schaeffer,
Nucl. Phys. A {\bf 635}, 231 (1998); {\bf 643}, 441(E) (1998).

\bibitem{Colo07}
G. Col\`{o}, H. Sagawa, S. Fracasso, and P. F. Bortignon, Phys.
Lett. B {\bf 646}, 227 (2007); {\bf 668}, 457(E) (2008).

\bibitem{Wang12}
M. Wang, G. Audi, A. H. Wapstra, F. G. Kondev, M. MacCormick, X.
Xu, and B. Pfeiffer, Chin. Phys. C {\bf 36}, 1603 (2012).

\bibitem{Birkhan16}
J. Birkhan, M. Miorelli, S. Bacca, S. Bassauer, C. A. Bertulani,
G. Hagen, H. Matsubara, P. von Neumann-Cosel, T. Papenbrock, N.
Pietralla, V. Yu. Ponomarev, A. Richter, A. Schwenk, and A. Tamii,
arXiv:1611.07072v2 [nucl-ex].

\bibitem{Piekarewicz12}
J. Piekarewicz, B. K. Agrawal, G. Col\`{o}, W. Nazarewicz, N.
Paar, P.-G. Reinhard, X. Roca-Maza, and D. Vretenar, Phys. Rev. C
{\bf 85}, 041302(R) (2012).

\bibitem{Hagen16}
G. Hagen, A. Ekstr\"{o}m, C. Forss\'{e}n, G. R. Jansen, W.
Nazarewicz, T. Papenbrock, K. A. Wendt, S. Bacca, N. Barnea, B.
Carlsson, C. Drischler, K. Hebeler, M. Hjorth-Jensen, M. Miorelli,
G. Orlandini, A. Schwenk, and J. Simonis, Nature Physics {\bf 12},
186 (2016).

\bibitem{Soloviev61}
V. G. Soloviev, Kgl. Dan. Vid. Selsk. Mat. Fys. Skr. {\bf 1}, 238
(1961).

\bibitem{Dobaczewski09}
J. Dobaczewski, W. Satu{\l}a, B. G. Carlsson, J. Engel, P.
Olbratowski, P. Powa{\l}owski, M. Sadziak, J. Sarich, N. Schunck,
A. Staszczak, M. Stoitsov, M. Zalewski, and H. Zdu\'{n}czuk,
Comput. Phys. Commun. {\bf 180}, 2361 (2009).

\bibitem{Grasso14}
M. Grasso, Phys. Rev. C {\bf 89}, 034316 (2014).

\bibitem{Holt12}
J. D. Holt, T. Otsuka, A. Schwenk, and T. Suzuki, J. Phys. G {\bf
39}, 085111 (2012).

\bibitem{Hagen12}
G. Hagen, M. Hjorth-Jensen, G. R. Jansen, R. Machleidt, and T.
Papenbrock, Phys. Rev. Lett. {\bf 109}, 032502 (2012).

\bibitem{Wu00}
S.-C. Wu, Nucl. Data Sheets {\bf 91}, 1 (2000).

\bibitem{Kibedi02}
T. Kib\'{e}di and R. H. Spear, At. Data Nucl. Data Tables {\bf
80}, 35 (2002).

\bibitem{Burrows06}
T. W. Burrows, Nucl. Data Sheets {\bf 107}, 1747 (2006).

\bibitem{Montanari12}
D. Montanari, S. Leoni, D. Mengoni, J. J. Valiente-Dobon, G.
Benzoni, N. Blasi, G. Bocchi, P. F. Bortignon, S. Bottoni, A.
Bracco, F. Camera, P. Casati, G. Col\`{o}, A. Corsi, F. C. L.
Crespi, B. Million, R. Nicolini, O.Wieland, D. Bazzacco, E.
Farnea, G. Germogli, A. Gottardo, S. M. Lenzi, S. Lunardi, R.
Menegazzo, G. Montagnoli, F. Recchia, F. Scarlassara, C. Ur, L.
Corradi, G. de Angelis, E. Fioretto, D. R. Napoli, R. Orlandi, E.
Sahin, A. M. Stefanini, R. P. Singh, A. Gadea, S. Szilner, M.
Kmiecik, A. Maj, W. Meczynski, A. Dewald, T. Pissulla, and G.
Pollarolo, Phys. Rev. C {\bf 85}, 044301 (2012).

\bibitem{Derya16}
V. Derya, N. Tsoneva, T. Aumann, M. Bhike, J. Endres, M. Gooden,
A. Hennig, J. Isaak, H. Lenske, B. L\"{o}her, N. Pietralla, D.
Savran, W. Tornow, V. Werner, and A. Zilges, Phys. Rev. C {\bf
93}, 034311 (2016).

\bibitem{Keefe87}
G. J. O'Keefe, M. N. Thompson, Y. I. Assafiri, R. E. Pywell, and
K. Shoda, Nucl. Phys. A {\bf 469}, 239 (1987).

\bibitem{Lipparini89}
E. Lipparini and S. Stringari, Phys. Rep. {\bf 175}, 103 (1989).

\bibitem{Tamii11}
A. Tamii, I. Poltoratska, P. von Neumann-Cosel, Y. Fujita, T.
Adachi, C. A. Bertulani, J. Carter, M. Dozono, H. Fujita, K.
Fujita, K. Hatanaka, D. Ishikawa, M. Itoh, T. Kawabata, Y.
Kalmykov, A. M. Krumbholz, E. Litvinova, H. Matsubara, K.
Nakanishi, R. Neveling, H. Okamura, H. J. Ong, B.
\"{O}zel-Tashenov, V. Yu. Ponomarev, A. Richter, B. Rubio, H.
Sakaguchi, Y. Sakemi, Y. Sasamoto, Y. Shimbara, Y. Shimizu, F. D.
Smit, T. Suzuki, Y. Tameshige, J. Wambach, R. Yamada, M. Yosoi,
and J. Zenihiro, Phys. Rev. Lett. {\bf 107}, 062502 (2011).

\bibitem{Bohigas81}
O. Bohigas, Nguyen Van Giai, and D. Vautherin, Phys. Lett. B {\bf
102}, 105 (1981).

\bibitem{Satula06}
W. Satu{\l}a, R. A. Wyss, and M. Rafalski, Phys. Rev. C {\bf 74},
011301(R) (2006).

\bibitem{Arsenyev10}
N. N. Arsenyev and A. P. Severyukhin, Phys. Part. Nucl. Lett. {\bf
7}, 112 (2010).

\bibitem{Vretenar01}
D. Vretenar, N. Paar, P. Ring, and G. A. Lalazissis, Nucl. Phys. A
{\bf 692}, 496 (2001).

\end{thebibliography}
\end{document}